\documentclass[man, 12pt, floatsintext]{apa7}
\usepackage[american]{babel}
\usepackage{amsmath}
\usepackage{amssymb}
\usepackage{setspace}
\usepackage{color}
\usepackage{enumitem}
\usepackage{verbatimbox}
\usepackage{caption}
\usepackage{mathtools}
\usepackage{multirow}
\usepackage{array}
\usepackage{xparse}
\usepackage{tabularx}
\usepackage{csquotes}
\usepackage{fancyhdr}
\usepackage{float}
\usepackage{url}
\usepackage[style=apa,sortcites=true,sorting=nyt,backend=biber,maxcitenames=2]{biblatex}
\usepackage[figuresright]{rotating}
\usepackage{array}
\usepackage{etoolbox}

\usepackage[utf8]{inputenc}

\DeclareDelimFormat[textcite]{finalnamedelim}{\finalnamedelim}

\DeclareMathAlphabet{\mathbf}{OT1}{cmr}{b}{n}

\definecolor{blue1}{RGB}{40,80,156}

\setlist[itemize, 1]{leftmargin=*, nosep}
\setlist[enumerate, 1]{leftmargin=*, topsep=1ex, itemsep=0ex,
  label={(\alph*)}, labelindent=0pt}

\def\ex{\mathbb{E}}

\def\t{^\top}
\def\indep{\mkern3mu\hbox{$\perp\mkern-11mu\perp$}\mkern3mu}
\def\XX{\mathbf{X}}
\def\xx{\mathbf{x}}
\def\YY{\mathbf{Y}}

\def\ZZ{\mathbf{Z}}

\def\xxi{\boldsymbol{\xi}}

\def\SSigma{\boldsymbol{\Sigma}}

\allowdisplaybreaks
\NewDocumentCommand{\evalat}{sO{\big}mm}{%
  \IfBooleanTF{#1}
   {\mleft. #3 \mright|_{#4}}
   {#3#2|_{#4}}%
}

\renewcommand*\finalnamedelim{\addspace\&\space}

\makeatletter
\newcommand{\thickhline}{%
    \noalign {\ifnum 0=`}\fi \hrule height 0.9pt
    \futurelet \reserved@a \@xhline
}
\newcolumntype{"}{@{\hskip\tabcolsep\vrule width 1pt\hskip\tabcolsep}}
\makeatother

\title{Multilevel Regression Discontinuity Models with Latent Variables}
\shorttitle{MLRD with Latent Variables}
\author{Monica Morell, Youngjin Han, Muwon Kwon, Youjin Sung,\break Yang Liu, Ji Seung Yang}
\affiliation{University of Maryland}
\authornote{The research reported here was supported, in whole or in part, by the
	Institute of Education Sciences, U.S. Department of Education,
	through grant R305D220030 to the University of Maryland. The
	opinions expressed are those of the authors and do not represent the
	views of the Institute or the U.S. Department of Education.}

\abstract{Regression discontinuity (RD) analysis with latent variables as introduced by \textcite{morell2025regression}, offers a useful augmentation of the conventional RD by incorporating measurement model. This approach is particularly relevant in education research, where noisy proxy (e.g., observed test score) of underlying latent construct is adopted for the running variable. This extension enables extrapolation of average treatment effect (ATE) away from the cutoff score and assessment of heterogeneous treatment effects. However, a key limitation of the original framework is its single-level structure, which does not account for the multilevel structure commonly found in education data—such as students nested within classrooms or schools. In this study, we extend the framework to multilevel contexts. We discuss models for both hierarchical RD design, where treatment is assigned at the cluster level, and multisite RD design, where treatment is assigned at the individual level within clusters. In both cases, multilevel measurement model is incorporated to describe the relationship between the latent running variable and observed indicators. Monte Carlo simulations demonstrate recovery of ATEs including extrapolated estimates beyond the cutoff given adequate cluster-level sample sizes. The study highlights the applicability of RD analysis with latent variables for broader use in educational research, without being restricted by the limitations of multilevel data.}

\begin{document}
	\maketitle
\vspace{1em}
\section{Introduction}
The regression discontinuity (\cite{ThistlewaiteCampbell1960}) design is often used in education research, as it allows for causal conclusions without the need for randomization. In the RD design, individuals are assigned to treatment conditions based on whether their scores on a running variable (also called an ``assignment variable'' or a ``forcing variable'') are above/below a specified cutoff value. The design identifies causal treatment effect for those with running variable (RV) score at the cutoff, which is referred to as the local average treatment effect (LATE; \cite{angrist1996}; \cite{Goldberger1972}; \cite{Rubin1977}).
\\
RD Designs naturally emerge in evaluation studies as programs or interventions that have built-in eligibility criteria. A common challenge in using RD designs in education research is the multilevel nature of the data, where individuals are nested within clusters (e.g., classes, schools, districts). The multilevel structure complicates statistical modeling because observations within clusters are not independent, violating a key assumption of many traditional parametric procedures. Consequently, the use of standard statistical models becomes inappropriate (\cite{Burstein1980}; \cite{Goldstein2011}; \cite{KreftDeLeeuw1998}; \cite{Raudenbush1993}; \cite{SnijdersBosker2012}). In response to these challenges, RD designs that incorporate multilevel modeling have gained increasing attention (e.g., \cite{bulus2021bound}; \cite{bulus2022minimum}; \cite{RhoadsDye2016}; \cite{schochet2009statistical}), which we henceforth refer to as multilevel RD.
\\
Meanwhile, RD analysis in education research often relies on aggregate measures of observed indicators, such as test scores, that serve as proxies for underlying latent constructs. For example, a high school exit exam score (\cite{Ou2010}) or a math test score (\cite{JacobLefgren2004a, NomiAllensworth2009}) is used as the RV to represent students' academic achievement in RD studies. In this work, we focus on cases where such measures serve as the RV, although they could also be adopted for the outcome variable or even covariates. In these cases, the RV provides a noisy measure of an underlying latent construct. Recognizing this, the potential for integrating LV modeling into RD analysis has been explored. An advancement in this regard is the work of \textcite{morell2025regression}, which incorporates an LV model that captures the relationship between observed indicators and the underlying construct.
\\
The underlying constructs are operationalized as LVs in LV modeling frameworks, with the relationship between the LV and its observed indicators (e.g., test scores) specified through what is commonly referred to as measurement model. In line with the RD terminology, the aggregates of observed indicators can be referred to as the observed running variable (ORV), while the corresponding latent construct is termed the latent running variable (LRV). The key aspect of the framework is to define the average treatment effect (ATE) conditional on the LRV, not only ORV, by delving into the relationship between the ORV and LRV. This approach addresses several limitations of traditional RD analysis, such as its inability to account for heterogeneous treatment effects and its limitations in extrapolating the ATE beyond the cutoff due to the lack of overlapping observations between the treatment and control group.
\\
The purpose of this study is to extend the latent RD framework, previously applied in a single-level context, to a multilevel context to address the common presence of multilevel data structures in education research. We integrate the latent RD framework with multilevel RD to extend its applicability. This allows the useful framework for broader use in educational research without being limited by the constraints of multilevel data.
\\
In the following sections, we first provide a brief overview of RD designs applicable to multilevel data. We then introduce our proposed models, which integrate multilevel RD models, originally suggested for observed variable RD analysis, with LV model. We discuss inferences related to the ATE and the additional utilities these models offer. Next, we describe the estimation algorithm employed, followed by a Monte Carlo simulation study. Finally, we offer suggestions for future research.
\vspace{1em}
\section{Regression Discontinuity Designs in Multilevel Settings}
Given the widespread presence of multilevel data structures, adopting an appropriate research design is crucial for assessing the causal impact of educational interventions. While multilevel random assignment is typically preferred for the purpose, it is not always feasible due to ethical concerns or intentional selection of specific sub-samples. In such cases, multilevel RD designs have been proposed as a strong alternative, providing a viable method for causal evaluation in these contexts (e.g., \cite{bulus2021bound}; \cite{schochet2009statistical}).
\\
When designing RD studies in multilevel settings, researchers must carefully consider the levels at which the RV and outcome variable are measured. In education research, it is common to assign entire clusters (e.g., schools or districts) to treatment and control groups, as this aligns with how policies and services are implemented at the organizational level. This results in a two-level RD design, where treatment is assigned at a higher level, but outcomes are measured at a lower level.
For instance, consider a policy in which schools with more than 85\% of students receiving free or reduced-price lunch are allocated additional resources, and the outcome of interest is student achievement. In this case, the RV (percentage of students on free/reduced lunch) is measured at the school level, whereas the outcome (student achievement) is measured at the individual level. This setup is typically referred to as a \emph{Hierarchical RD design} (\cite{RhoadsDye2016}; also known as a Clustered RD Design, \cite{schochet2009statistical, bulus2022minimum}). Such designs are commonly seen in educational studies, where interventions often target entire schools or districts rather than individual students.
\\
On the other hand, a different scenario arises when both the RV and the outcome variable are measured at the individual level, but the sampling process is inherently clustered. For example, consider a statewide tutoring program aimed at students who score below a certain threshold on a standardized test. If certain schools are sampled for participation in the program, even though both the RV (test score) and the outcome (student performance after tutoring) are measured at the individual level, the clustering of the samples by school introduces a nested structure. Therefore, multilevel statistical models should be utilized that appropriately account for non-independent data. Such cases are examples of \emph{Multisite RD designs} (\cite{RhoadsDye2016}). In both designs, addressing the multilevel structure is critical for ensuring accurate estimation of the treatment effects and for drawing valid causal inferences from the study.
\vspace{1em}
\section{Multilevel Latent Regression Discontinuity Models}
We introduce multilevel latent RD models and associated inferences regarding the ATEs. The latent RD framework (\cite{morell2025regression}) is extended by incorporating multilevel RD designs. We consider two multilevel RD designs described in the previous section---Hierarchical RD and Multisite RD designs (\cite{RhoadsDye2016})---which can be utilized in multilevel settings for different scenarios of treatment assignment. The definition of ATEs conditional on the LRV not only on the ORV will be described, as a key feature allowed by the proposed model. Finally, extrapolation of ATEs apart from the cutoff and quantification of heterogeneous treatment effects will be discussed.

\subsection{Measurement Model} 
Let $i$ index individuals and $j$ index clusters: Individuals are nested within clusters, such that $j = 1,\dots, J$ and $i = 1,\dots, I_j$. For each individual $i$ within cluster $j$, let $\theta_{ij}$ be the LRV, measured by a collection of $K$ observed indicators $\XX_{ij} = (X_{ijk}: k = 1,\dots,K)$. In particular, it is assumed that
	\begin{align}
	\theta_{ij} = \theta_j + \delta_{ij},
	\label{eq:theta}
	\end{align}
\noindent in which $\theta_j$ denotes the cluster-level LRV (i.e., level-$2$ LRV) and $\delta_{ij}$ denotes the individual deviation around $\theta_j$ (i.e., level-$1$ LRV). $\theta_{j}$ is distributed as $\mathcal{N}(0, \psi^2)$ and $\delta_{ij}$ follows a standard normal distribution. 
\\
Pooling across all $i$'s within cluster $j$, let $\XX_j = (\XX_{ij}: i = 1,\dots,I_j)$. A measurement model specifies the likelihood of $\XX_j$ conditional on $\theta_{ij}$:
	\begin{align}
	f_{j}(\xx_j|\theta_{ij}) = \prod_{i=1}^{I_j}\prod_{k=1}^K
	f_{k}(x_{ijk}|\theta_{ij})
	\label{eq:mlikx}
	\end{align}
 
	\noindent in which $\xx_j = (x_{ijk}:i=1,\dots,I_j; k=1,\dots,K)$ denotes a realization of $\XX_{j}$. The double product in Equation \ref{eq:mlikx} results from the assumption of independence among $X_{ij1},\dots, X_{ijK}$ conditional on $\theta_{ij}$---a key assumption of factor analytic models (\cite{McDonald1981}). While the conditional likelihood of $x_{j}$ given $\theta_{ij}$ can take various functional forms, we focus on the following two-parameter logistic (2PL; \cite{Birnbaum1968}) model for dichotomous indicators.
\begin{align}
f_{k}(x_{ijk}|\theta_{ij}) = \frac{\exp[x_{ijk}(a_k\theta_{ij} +
	c_{k})]}{1 + \exp(a_k\theta_{ij} + c_{k})},
\label{eq:clikgrm}
\end{align}

\noindent in which  $a_k$ and $c_k$ denote the slope and intercept parameters, which are assumed to be invariant across individual- and cluster-level. 
Note we assume that the measurement model for the LRV is correctly specified, ensuring that the ATE in relation to the LRV is meaningful.

\subsection{Structural Model}
\noindent\textbf{Hierarchical RD}. In the HRD design, clusters themselves are assigned to different treatment conditions (e.g., all students in a school receive the intervention and all students in another school receive the control condition). Therefore, the treatment indicator is denoted  $T_j$ without the subscript $i$. In the current work, we consider a scenario in which all individuals within a cluster are assigned to treatment if their cluster-average ORV falls below a predetermined cutoff. Specifically, the ORV is calculated as the unweighted sum of all observed indicators used to measure the LRV. The treatment indicator is then defined as
\begin{align}
T_j =&
\begin{cases}
0 & \text{if } S(\boldsymbol{X}_j) > c, \\
1 & \text{if } S(\boldsymbol{X}_j) \leq c,
\end{cases}
\\
\quad \text{where } 
&S(\boldsymbol{X}_j) = \frac{1}{I_j} \sum_{i=1}^{I_j} \sum_{k=1}^{K} X_{ijk}.
\label{eq:treatment.hrd}
\end{align}
Under standard multilevel measurement models (e.g., the 2PL model), a cluster can be assigned to either the treatment or control condition with a strictly positive probability for any given LRV value:
 \begin{align}
    P\{T_{j} = 1|\theta_{j}=\theta\} \in (0,1).     
    \label{eq:hrd.positivity}
 \end{align}
 This is often referred to as the positivity assumption in RD analysis. 
 
 We constrain our HRD model to have random intercept but no random slopes following \textcite{RhoadsDye2016}'s model specification. The HRD model we consider is specified as follows.
 \begin{align}
	\text{Level 1:		}Y_{ij} &= \beta_{0j} + \beta_{1j}\delta_{ij} + \varepsilon_{ij}, \\ \nonumber
	\text{Level 2:		}\beta_{0j} &= \gamma_{00} + \gamma_{01}\theta_{j} + \gamma_{02}T_{j} + \gamma_{03} \theta_{j} T_j + u_{0j}, \\ \nonumber
	\beta_{1j} &= \gamma_{10}, \\ \nonumber
    \varepsilon_{ij} &\sim \mathcal{N}(0, \sigma^2) \;\;\text{and}\;\; u_{0j} \sim \mathcal{N}(0, \tau_{0}^2).
	\end{align}

\noindent in which $Y_{ij}$ is the outcome variable, $\theta_{j}$ represents the cluster-level LRV and $\delta_{ij}$ represents the individual-level deviation from $\theta_{j}$, which, per convention in IRT, each follows normal distribution. $\gamma_{00}$ is the intercept. $\gamma_{01}$, $\gamma_{02}$ and $\gamma_{10}$ represent the partial effects of the cluster-level LRV, treatment, and individual-level LRV, respectively. $\gamma_{03}$ is the interaction effect between the LRV and treatment. 
\\
The equation can be also written using the potential outcomes notation as
\begin{align}
    Y_{ij} =&\ T_{j}Y_{ij}(1) + (1-T_{j})Y_{ij}(0), \cr
    Y_{ij}(1) =&\ (\gamma_{00} + \gamma_{02}) + (\gamma_{01} + \gamma_{03})\theta_{j} + \gamma_{10}\delta_{ij} + u_{0j} + \varepsilon_{ij}(1), \cr
    Y_{ij}(0) =&\ \gamma_{00} + \gamma_{01}\theta_{j} + \gamma_{10}\delta_{ij} + u_{0j} + \varepsilon_{ij}(0),
    \label{eq:HRD.potential.outcome}
\end{align}
\noindent where $Y_{ij}(1)$ is the potential outcome for individual $i$ in cluster $j$ under treatment condition and $Y_{ij}(0)$ is that under control condition. $\varepsilon_{ij}(1)$ and $\varepsilon_{ij}(0)$ are assumed to be independent and normally distributed with mean zero and variance $\sigma^2$. The equivalent combined equation is
	\begin{equation}
	Y_{ij}(g) = \gamma_{00} + \gamma_{01}\theta_{j} + \gamma_{02}g + \gamma_{03} \theta_{j} g  + \gamma_{10}\delta_{ij} + u_{0j} + \varepsilon_{ij}(g), g=0, 1.
    \label{eq:HRD.comb}
\end{equation}
 From Equation \ref{eq:HRD.comb} and the normality of random effects, the probability density function of the HRD structural model can be written as 
    \begin{align}
        f_{y(g)|\delta, \theta, u}(y_{ij}|\delta_{ij}, & \theta_{j}, u_{0j}) = \cr
        \frac{1}{\sqrt{2\pi\sigma^2}} \exp &\left(-\frac{(y_{ij} - \gamma_{00} - \gamma_{01}\theta_{j} - \gamma_{02}g - \gamma_{03} \theta_{j} g  - \gamma_{10}\delta_{ij} - u_{0j})^2}{2\sigma^2} \right), g=0, 1.
    \end{align}
    
\noindent \textbf{Multisite RD}. In the MRD design, the assignment of the treatment occurs at the individual level. Individuals within the same cluster can be assigned to either treatment or control group based on an individual-level variable. Since individuals still exist in clusters, they are not independent of each other, and therefore multilevel modeling is required.
In the current work, we consider individual's summed score as an ORV that determines the treatment status as follows: 
\begin{align}
T_{ij} =&
\begin{cases}
0 & \text{if } S(\boldsymbol{X}_{ij}) > c, \\
1 & \text{if } S(\boldsymbol{X}_{ij}) \leq c,
\end{cases}
\quad \cr
\text{where } 
&S(\boldsymbol{X}_{ij}) = \sum_{k=1}^{K} X_{ijk}.
\label{eq:treatment.mrd}
\end{align}
Note that the cutoff $c$ is not allowed to vary by cluster.
Similar to the HRD model, it is assumed that an individual can be assigned with a positive probability to either the treatment or control group given any individual LRV value: that is,
 \begin{align}
    P\{T_{ij} = 1|\delta_{ij}=\delta\} \in (0,1).
    \label{eq:mrd.positivity}
 \end{align}
For the MRD model, we consider two random effects---one for the intercept and the other for the treatment effect. The structural part of the MRD model is then

\begin{align}
        \text{Level 1:		}Y_{ij} &= \beta_{0j} + \beta_{1j}\delta_{ij} + \beta_{2j}T_{ij} + \beta_{3j}\delta_{ij}T_{ij} + \varepsilon_{ij}  \\ \nonumber
        \text{Level 2:		}\beta_{0j} &= \gamma_{00} + \gamma_{01}\theta_{j} + u_{0j} \\ \nonumber
        \beta_{1j} &= \gamma_{10} \\ \nonumber
        \beta_{2j} &= \gamma_{20} + u_{2j} \\ \nonumber
        \beta_{3j} &= \gamma_{30} \\ \nonumber
    \varepsilon_{ij} &\sim \mathcal{N}(0, \sigma^2) \;\;\text{and}\;\; (u_{0j}, u_{2j})\t \sim \mathcal{N}\left(\boldsymbol{0}, \; \begin{pmatrix}
	\tau_0^2 & \tau_{02}\\
	\tau_{02} & \tau_2^2\\
	\end{pmatrix}
    \right).
    \end{align}
\noindent where $Y_{ij}$ is the outcome variable, $\theta_{j}$ and $\delta_{ij}$ are the normally distributed cluster- and individual-level LRV. $\gamma_{00}$ is the intercept, $\gamma_{01}$, $\gamma_{10}$ and $\gamma_{20}$ are the effect of the cluster-level LRV, individual-level LRV, and treatment, respectively. $\gamma_{30}$ is the interaction effect between the individual-level LRV and treatment. The individual-level random effects, $\varepsilon_{ij}$ are distributed as $\mathcal{N}(0, \sigma^2)$. $u_{0j}$ and $u_{2j}$ are cluster-level random effects for the intercept and treatment effect, respectively. They jointly follow a bivariate normal distribution.    
\\
Similar to the HRD model, the potential outcomes from the MRD model can be written as
\begin{align}
    Y_{ij}(1) =&\ (\gamma_{00} + \gamma_{20} + u_{2j}) + \gamma_{01}\theta_{j} + (\gamma_{10} + \gamma_{30})\delta_{ij} + u_{0j} + \varepsilon_{ij}(1), \cr
    Y_{ij}(0) =&\ \gamma_{00} + \gamma_{01}\theta_{j} + \gamma_{10} \delta_{ij} + u_{0j} + \varepsilon_{ij}(0).
    \label{eq:MRD.potential.outcome}
\end{align}
\noindent The equivalent combined equation is
    \begin{equation}
        Y_{ij}(g) = \gamma_{00} + \gamma_{01}\theta_{j} + \gamma_{10}\delta_{ij} + (\gamma_{20} + u_{2j})g + \gamma_{30}\delta_{ij}g + u_{0j} + \varepsilon_{ij}(g), g=0, 1, 
    \label{eq:MRD.comb}
    \end{equation}
From Equation \ref{eq:MRD.comb} and the normality of random effects, the probability density function of the MRD structural model can be written as
    \begin{align}
        f_{y(g)|\delta, \theta, \boldsymbol{u}}(y_{ij}|\delta_{ij}, & \theta_{j}, u_{0j}, u_{2j}) = \cr
        \frac{1}{\sqrt{2\pi\sigma^2}} \exp &\left(-\frac{[y_{ij} - \gamma_{00} - \gamma_{01}\theta_{j} - \gamma_{10}\delta_{ij} - (\gamma_{20} + u_{2j})g - \gamma_{30} \delta_{ij} g - u_{0j}]^2}{2\sigma^2} \right).
    \end{align}   

\subsection{Inferences}
\textbf{Average Treatment Effects.}
A key aspect of the latent RD framework is its capability to identify ATEs conditioning not only on the ORV but also the LRV. The treatment effect from the conventional RD design is defined as $\ex[Y_{i}(1) - Y_{i}(0)]$ (\cite{Rubin1974}), where $Y_{i}(1)$ is the potential outcome of individual $i$ under treatment condition and $Y_{i}(0)$ is that under control condition. The following equations present the ATEs identifiable from the proposed latent RD model for HRD and MRD designs, respectively. 
\\
Using Equation \ref{eq:HRD.potential.outcome}, the ATEs conditional on LRV under the HRD model can be defined as
\begin{align}
    \omega_{HRD}(\theta_j) &= 
    \ex\left[Y_{ij}(1) - Y_{ij}(0) \,|\, \theta_j, \delta_{ij}, u_{0j} \right] \cr
    &= \ex\left[Y_{ij}(1) \,|\, \theta_j, \delta_{ij}, u_{0j} \right] - \ex\left[Y_{ij}(0) \,|\, \theta_j, \delta_{ij}, u_{0j} \right] \cr
    &= \gamma_{01} + \gamma_{03}\theta_j
    \label{eq:omega.hrd}
\end{align}

\noindent Then, the ATEs conditional on an ORV score can be derived as follows.
\begin{align}
    \tau_{HRD}(s) &= \ex\left[Y_{ij}(1) - Y_{ij}(0) \mid S(\boldsymbol{X}_j)=s\right] \cr
    &= \ex\left[\ex \left\{ Y_{ij}(1) - Y_{ij}(0) \mid \theta_{j}, S(\boldsymbol{X}_j)=s \right\} \mid S(\boldsymbol{X}_j)=s\right] \cr
    &= \ex\left[ \omega_{HRD}(\theta_j) \mid S(\boldsymbol{X}_j)=s \right].
    \label{eq:tau.hrd}
\end{align}
To establish the last equality of Equation \ref{eq:tau.hrd}, we rely on the assumption that the potential outcomes are independent of the ORV conditional on the LRV and random effects.
\begin{align}
    \left( Y_{ij}(1), Y_{ij}(0) \right) \indep S(\boldsymbol{X}_{j} = s) \,|\, \theta_{j}, \delta_{ij}, u_{0j} 
    \label{eq:indep.multilevel}
\end{align}
Combining Equations \ref{eq:omega.hrd}, \ref{eq:tau.hrd} and \ref{eq:indep.multilevel}, we obtain
\begin{align}
    \tau_{HRD}(s) = \gamma_{01} + \gamma_{03}\ex(\theta_{j}|S(\boldsymbol{X}_j)=s).
    \label{eq:tau.hrd2}
\end{align}
We see from Equation \ref{eq:tau.hrd2} that the ATE conditional on ORV can be interpreted as the average of $\omega_{HRD}(\theta_j)$ among whose ORV score is $s$. Also, note that the ATEs from the HRD model is the treatment effect defined at the cluster-level.
\\
In a similar way, we can define ATEs under the MRD model. We start with the ATE conditional on all latent variables and random effects. By Equation \ref{eq:MRD.potential.outcome}, let
\begin{align}
    \omega_{MRD}(u_{2j}, \delta_{ij}) &= 
    \ex\left[Y_{ij}(1) - Y_{ij}(0) \,|\, \theta_j, \delta_{ij}, u_{0j}, u_{2j} \right] \cr
    &= \ex\left[Y_{ij}(1) \,|\, \theta_j, \delta_{ij}, u_{0j}, u_{2j} \right] - \ex\left[Y_{ij}(0) \,|\, \theta_j, \delta_{ij}, u_{0j}, u_{2j} \right] \cr
    &= \gamma_{20} + u_{2j} + \gamma_{30}\delta_{ij}.
    \label{eq:omega.mrd}
\end{align}
Using Equation \ref{eq:omega.mrd}, we can define the cluster-specific and overall ATEs conditional on the ORV score level $s$. For a specific cluster $j$, the ATE conditional on an ORV score $s$ can be expressed as
\begin{align}
    \tau_{MRD}(s, u_{2j},\theta_j) 
    &= \ex\left[Y_{ij}(1)-Y_{ij}(0) \mid S(\boldsymbol{X}_{ij})=s, \theta_{j}, u_{0j}, u_{2j} \right] \cr
    &= \ex \left[ \ex\left\{Y_{ij}(1)-Y_{ij}(0) \mid \delta_{ij}, S(\boldsymbol{X}_{ij})=s, \theta_{j}, u_{0j}, u_{2j} \right\} \mid S(\boldsymbol{X}_{ij})=s, \theta_{j}, u_{0j}, u_{2j} \right] \cr
    &= \ex \left[ \omega_{MRD}(u_{2j}, \delta_{ij}) \mid S(\boldsymbol{X}_{ij})=s, \theta_{j}, u_{0j}, u_{2j} \right]    \cr
    &= \gamma_{20}+u_{2j}+\gamma_{30}\,\ex\left(\delta_{ij} \mid S(\boldsymbol{X}_{ij})=s, \theta_j \right),
    \label{eq:tau.mrdj}
\end{align}
in which the last equality follows from the fact that $u_{0j}$ and $u_{2j}$ are independent of $\theta_j$, $\delta_{ij}$, and $S(\boldsymbol{X}_{ij})$.
Next, we define the overall ATE conditional on an ORV score $s$ by
\begin{align}
    \tau_{MRD}(s) &= \ex\left[Y_{ij}(1) - Y_{ij}(0) \mid S(\boldsymbol{X}_{ij})=s\right] \cr
    &= \ex\left[\ex \left\{ Y_{ij}(1) - Y_{ij}(0) \mid \theta_j, u_{0j}, u_{2j}, S(\boldsymbol{X}_{ij})=s \right\} \mid S(\boldsymbol{X}_{ij})=s\right] \cr
    &= \ex\left[ \omega_{MRD}(u_{2j}, \delta_{ij}) \mid S(\boldsymbol{X}_{ij})=s  \right] \cr
    &= \gamma_{20} + \gamma_{30}\ex(\delta_{ij} \mid S(\boldsymbol{X}_{ij})=s).
    \label{eq:tau.mrd}
\end{align}
\noindent Note that $\tau_{MRD}(s)$ represents an individual-level treatment effect regardless of the cluster membership. Under the current model specification, the ATE conditional on the LRVs (Equation \ref{eq:omega.mrd}) incorporates cluster-level variability solely through $u_{2j}$. However, because $\ex[u_{2j} \mid S(\boldsymbol{X}_{ij})=s]=0$, $\tau_{MRD}(s)$, does not depend on any cluster-level variability.

\begin{figure}[H]
    \begin{center}
	\includegraphics[width=13cm]{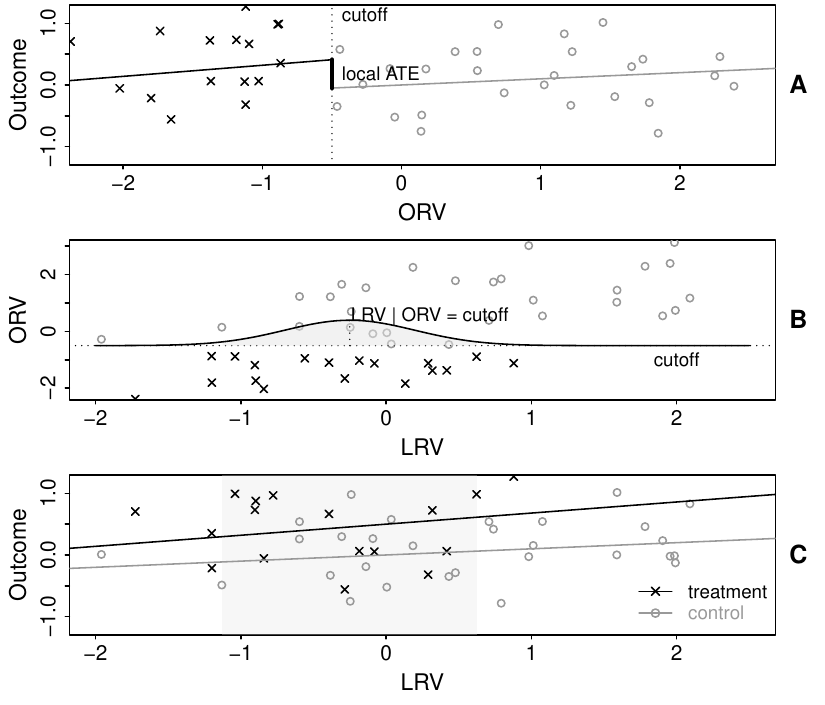}
    \end{center}
\caption{Illustrative example of RD analysis. Panel A: Outcome plotted against ORV. The ATE is only identified at the cutoff ($c=-0.5$). 
Panel B: Posterior density of LRV given ORV. The variability quantifies the imperfect correspondence between LRV and ORV due to measurement error. Shaded area represents middle $95\%$ of the density. 
Panel C: Outcome plotted against LRV. The overlap between the treatment and control group can be claimed not only at the cutoff.}
\label{fig:rd.illust}
\end{figure}

\noindent \textbf{Extrapolation of Average Treatment Effect.}
One limitation of conventional RD analysis is that the ATE is only identified at the cutoff. This is because the design inherently assumes that each observation belongs exclusively to either the treatment or control group, but not both. Consequently, potential outcomes for both the treatment and control groups are inferable only at the cutoff, using the limiting properties of continuous functions at a point (Figure \ref{fig:rd.illust}A). However, stakeholders often seek to evaluate the effectiveness of a program across a wider range of ORV scores, beyond just the cutoff. This helps inform a comprehensive evaluation of the program and supports decision-making, such as adjusting the cutoff for future study.
\\

Latent RD framework allows extrapolating the ATE conditional on arbitrary ORV away from the cutoff. This is due to the imperfect correspondence between ORV and LRV, which is also commonly referred to as measurement error. As a result of this measurement error, we can assume a positive probability that an individual may be assigned to either the treatment or control group at any given LRV value (Figure \ref{fig:rd.illust}B). This justifies fitting the regression model for both groups, across entire domain, as well as computing the ATE given LRV at any given value (Figure \ref{fig:rd.illust}C). As $\tau(s)$ is computed as a conditional average of $\omega$, computing $\tau(s)$ at an arbitrary value of $s$ is also justified.
\\
\noindent \textbf{Heterogeneity of Treatment Effects.}
Within the conventional RD analysis, the LATE is interpreted as a homogeneous treatment effect for the subset of individuals whose RV is at the cutoff. On the other hand, we see from the Equation \ref{eq:tau.hrd} that the ATE conditional on ORV can be seen as the weighted average of $\omega_{HRD}(\theta_j)$ among whose ORV score is $s$. This equation also implies that the heterogeneous treatment effect can be quantified, which is dependent on the variability of the posterior distribution $\omega \mid s$. The variance of the posterior distribution can be written as
\begin{align}
    \mathrm{Var}[\omega_{HRD}(\theta_j) \mid S(\boldsymbol{X}_j)=s] = \gamma_{03}^2 \mathrm{Var}[\theta_{j} \mid S(\boldsymbol{X}_j)=s]
\end{align}
\label{eq:hrd.var.post}
for the HRD model and
\begin{align}
    \mathrm{Var}[\omega_{MRD}(u_{2j}, \delta_{ij}) \mid S(\boldsymbol{X}_{ij})=s] = \mathrm{Var}(u_{2j}) + \gamma_{30}^2 \mathrm{Var}[\delta_{ij} \mid S(\boldsymbol{X}_{ij})=s]
    \label{eq:mrd.var.post}
\end{align}
for the MRD model. We can also quantify the $\alpha$th quantile of the ATE among those who share the same ORV. That can be computed as
\begin{align}
    Q_{\alpha}[\omega_{HRD}(\theta_j) \mid S(\boldsymbol{X}_j)=s] = \gamma_{01} + \gamma_{03} Q_{\alpha}[\theta_j \mid S(\boldsymbol{X}_j)=s],
    \label{eq:hrd.hetero}
\end{align}
for the HRD model and
\begin{align}
    Q_{\alpha}[\omega_{MRD}(u_{2j}, \delta_{ij}) \mid S(\boldsymbol{X}_{ij})=s] = \gamma_{20} + Q_{\alpha}[u_{2j} + \gamma_{30}\delta_{ij} \mid S(\boldsymbol{X}_{ij})=s].
    \label{eq:mrd.hetero}
\end{align}
for the MRD model, where $Q_{\alpha}[\cdot]$ represents a general notation for the $\alpha$th quantile of a given distribution. For instance, the quantity computed by Equation \ref{eq:hrd.hetero} can be interpreted as $\alpha$th quantile of the ATE among schools where their school average ORV is at $s$. The shaded area in Figure \ref{fig:rd.illust} visualizes the variability of LRV given ORV. In Panel C, varying discrepancies between the two lines within the shaded area shows the heterogeneous treatment effect among observations whose ORV is $-0.5$.
\\
For the MRD model, the quantification of heterogeneity by Equations \ref{eq:mrd.var.post} and \ref{eq:mrd.hetero} is defined at individual level regardless of the cluster membership. In multilevel analyses, however, within-cluster heterogeneity could be of greater interest. This can be quantified by using the posterior distribution $\omega_{MRD} \mid S(\boldsymbol{X}_{ij})=s, \theta_j,  u_{0j}, u_{2j}$. The variance and quantile of this distribution can be written as
\begin{align}
   \mathrm{Var} [\omega_{MRD}(u_{2j}, \delta_{ij}) \mid S(\boldsymbol{X}_{ij})=s, \theta_j, u_{0j}, u_{2j}]  = \gamma_{30}^2 \mathrm{Var}[\delta_{ij} \mid S(\boldsymbol{X}_{ij})=s, \theta_j]
    \label{eq:mrd.var.post.within}
\end{align} 
and
\begin{align}
    Q_{\alpha}[\omega_{MRD}(u_{2j}, \delta_{ij}) \mid S(\boldsymbol{X}_{ij})=s, \theta_j, u_{0j}, u_{2j} ] = \gamma_{20} + u_{2j} + \gamma_{30}Q_{\alpha}[\delta_{ij} \mid S(\boldsymbol{X}_{ij})=s, \theta_j].
    \label{eq:mrd.hetero.within}
\end{align}
\subsection{The MH-RM Algorithm for Estimation}
In this section, we briefly describe the Metropolis-Hastings Robbins-Monro (MH-RM; \cite{Cai2010a}; \cite{Cai2010b}) algorithm we use for the parameter estimation. The MH-RM algorithm is a stochastic approximation based variant of the Newton-Raphson type algorithm, which approximates the observed data likelihood by using Fisher's identity for data augmentation via a Metropolis-Hastings (MH; \cite{Hastings1970}, \cite{MetropolisEtAl1953}) sampler and then employs the Robbins-Monro (RM; \cite{RobbinsMonro1951}) stochastic approximation for parameter updates. Because the MH-RM algorithm does not use numerical integration, it is suitable for estimating models with many LVs.
\\
The HRD and MRD models with LV involve three and four LVs/random effects, respectively, and therefore, the computational efficiency improves by using the MH-RM compared to the quadrature-based methods. In addition, the current work constrained the model by assuming that there is no covariate and the outcome variable is observed. However, the model can be extended to have latent outcome variable and covariates as well. Given the MH-RM algorithm, such extensions are feasible with minimal concern for computational burden.
\\
Each iteration, $t = 1,\dots, T$, of the MH-RM algorithm consists of three steps: Stochastic Imputation, Stochastic Approximation, and Robbins-Monro Update.\\
\textbf{Step 1: Stochastic imputation}.
Let $f$ be the general symbol for probability density/mass functions and $\xxi$ be all the parameters in the measurement and structural models. The ML estimator of $\xxi$ can be obtained by maximizing the log-likelihood function $\log f(\XX, \YY|\xxi)$, where $\XX$ is the observed indicator that measures the LRV and $\YY$ is the observed outcome variable. Let $\mathbf{Z}$ collect all missing data that includes individual- and cluster-level LRV, random intercept, and random slope. These random effects are drawn from a Markov chain that targets the posterior predictive distribution of missing data ($\ZZ$) given the observed data $(\mathbf{X}, \mathbf{Y})$, i.e., $f(\mathbf{Z}|\XX, \YY, \xxi^{(t)})$, where $\xxi^{(t)}$ is the current estimates of model parameters at iteration $t:t = 1,\dots, T$.   At each iteration, $M_t$ sets of complete data are formed as follows: \\
\begin{equation}
\{\mathbf{Y}, \mathbf{X}, \mathbf{Z}_m^{(t+1)}; m = 1, \dots, M_t\}
\end{equation}\\
\textbf{Step 2: Stochastic approximation}.
The gradient vector of the complete data log-likelihood function is defined as:\\ 
\begin{equation}
\boldsymbol{s}(\boldsymbol{\xi}| \boldsymbol{Y}, \boldsymbol{X}, \boldsymbol{Z}) = \frac{\partial}{\partial\boldsymbol{\xi}}\log f(\boldsymbol{Y},
\boldsymbol{X}, \boldsymbol{Z}|\boldsymbol{\xi}).
\end{equation}\\
By Fisher's \citeyear{Fisher1925} identity, the gradient of the observed data log-likelihood is the expectation of $\boldsymbol{s}(\xxi| \XX, \YY, \ZZ)$ over the posterior distribution $f(\mathbf{Z|X,Y},\xxi)$: that is,\\
\begin{equation}
\frac{\partial}{\partial \xxi}\log f(\XX, \YY|\xxi) = \int\boldsymbol{s}(\xxi| \XX, \YY, \ZZ) f(\ZZ|\XX, \YY, \xxi)d\ZZ.
\label{eq:grad}
\end{equation}\\
Equation \ref{eq:grad} evaluated at $\xxi^{(t)}$ gives the direction of steepest ascent. Direct evaluation of the observed data gradient is computationally expensive when the integration is high-dimensional. Nevertheless, given a
set of imputed missing data $\ZZ_m^{(t+1)}$, $m = 1,\dots, M_t$, Equation \ref{eq:mc.grad} suggests the Monte Carlo estimates of the observed data gradient:

\begin{equation}
\tilde{\boldsymbol{s}}_{t+1} = \frac{1}{m_t}\sum_{m=1}^{M_t} \boldsymbol{s}(\boldsymbol{\xi}^{(t)}|
\boldsymbol{Y}, \boldsymbol{X}, \boldsymbol{Z}_m^{(t+1)}).
\label{eq:mc.grad}
\end{equation}
\\
\noindent $\tilde{\boldsymbol{s}}_{t+1}$ gives a noise-corrupted version of $s(\xxi^{(t)}|\boldsymbol{X}, \boldsymbol{Y})$. \\
\textbf{Step 3: Robbins-Monro update}.
The following equation suggests an approximation of the conditional expectation of the complete-data information matrix at the $(t+1)th$ iteration (e.g., \cite{Cai2008}, \cite{GuKong1998}).\\
\begin{equation}
\boldsymbol{\Gamma}_{t+1} = \boldsymbol{\Gamma}_t + g_t\left[\frac{1}{M_t} \sum_{m=1}^{M_t}
\boldsymbol{H}(\boldsymbol{\xi}^{(t)} | \XX, \YY, \boldsymbol{Z}_m^{(t+1)}) - \boldsymbol{\Gamma}_t\right],
\end{equation}\\
\noindent where the complete data information matrix is\\
\begin{equation}
\boldsymbol{H}(\boldsymbol{\xi}|\XX, \YY, \boldsymbol{Z}) = -\frac{\partial^2}{\partial
	\boldsymbol{\xi} \partial 
	\boldsymbol{\xi}^{'}} \log f(\boldsymbol{\xi}|\XX, \YY,\boldsymbol{Z}).
\end{equation}\\
\noindent Then parameters are updated recursively as\\
\begin{equation}
\boldsymbol{\xi}^{(t+1)} = \boldsymbol{\xi}^{(t)} + g_t\boldsymbol{\Gamma}_{t+1}^{-1}
\tilde{\boldsymbol{s}}_{t+1}.
\end{equation}\\
\noindent Details for the first and second derivatives of the complete data log-likelihood can be found in the Appendix. ${g_t; t >1}$ is a decaying sequence of gain constants, which can be defined to filter out noise across iterations. In practice, different gain constants are used in different stages of the algorithm. A single gain constant value is used across the initial burn-in stage (i.e., Stage I) iterations. The second stage (i.e., Stage II) also adopts a single gain constant value in all iterations with the goal of obtaining good starting values for the next stage. The third stage (i.e., Stage III) often starts with the average values of the Stage II estimates and employs a decaying gain sequence that satisfies $\sum_{t=1}^\infty g_t = \infty$ and $\sum_{t=1}^\infty g_t^2 < \infty$ to reduce the impact of Monte Carlo noise. The algorithm is terminated once the minimum change in a parameter estimate is below a desired threshold for a window of iterations (\cite{Cai2008}).
\\
\textbf{Step 4: Standard error estimation.}
Standard errors are estimated after the convergence of the algorithm. From the Louis formula (\cite{Louis1982}), the observed information matrix is
\begin{align} \label{eq:sss:se}
\mathcal{I} &= E_{\ZZ} \{ \boldsymbol{H}(\boldsymbol{X}, \boldsymbol{Y}, \boldsymbol{Z}; \boldsymbol{\xxi})|\boldsymbol{X}, \boldsymbol{Y}\} \\ \nonumber
&- E_{\ZZ}\{\boldsymbol{s}(\boldsymbol{X}, \boldsymbol{Y}, \boldsymbol{Z}; \boldsymbol{\xxi})\boldsymbol{s}^{\top} (\boldsymbol{X}, \boldsymbol{Y}, \boldsymbol{Z}; \boldsymbol{\xxi})|\boldsymbol{X}, \boldsymbol{Y} \}, \label{eq:louis} \\ 
&+ E_{\ZZ}\{\boldsymbol{s}(\boldsymbol{X}, \boldsymbol{Y}, \boldsymbol{Z}; \boldsymbol{\xxi})|\boldsymbol{X}, \boldsymbol{Y} \}  E_{\ZZ}\{\boldsymbol{s}^{\top}(\boldsymbol{X}, \boldsymbol{Y}, \boldsymbol{Z}; \boldsymbol{\xxi})|\boldsymbol{X}, \boldsymbol{Y} \}, \nonumber
\end{align}
\noindent The expectations over $\ZZ|\XX, \YY$ are often approximated by Monte-Carlo imputed sample averages. That is, we draw $M$ samples of $\tilde{\ZZ}$ from the posterior distribution $f(\ZZ|\XX, \YY)$ and get the average to approximate components in Equation \ref{eq:sss:se}. Usually, a number greater than $1,000$ is recommended for $M$ (\cite{flexmirt}).
\vspace{1em}
\section{Simulation study}
A Monte-Carlo simulation study was conducted to verify the proposed model. We examine the recovery of the parameters as well as the recovery of the ATE computed as a combination of the parameter estimates. In addition, we demonstrate inferences related to ATE, including the computation of ATE away from the cutoff and quantification of the heterogeneous ATE given certain ORV value.
\subsection{Data generation}
In this simulation study, we manipulate the cluster-level sample size ($J = 100, 300$, and $500$), as our preliminary analysis indicates that the cluster-level sample size significantly influences the recovery of parameters, particularly the interaction coefficient and variance components. The clusters were balanced with each having $I_{j}=30 \, (j=1, \dots, J)$. Individual-level units within the cluster, which was chosen based on typical sample sizes in education research assessing classroom or school traits (\cite{ludtke2008multilevel}).

Data generation comprises three steps. \\
\noindent\textbf{Step 1: Generating item parameters and latent variables} The length of the test that measures the LRV was fixed at $10\,(K=10)$. The 2PL model as described in Equation \ref{eq:clikgrm} is the data generating model. The slope parameters ($a_{k}$) were drawn from a lognormal distribution with a mean of $0.3$ and a standard deviation of $0.20$, truncated to the interval [1.0, 2.5]. The difficulty parameters ($b_{k}$) were generated from a standard normal distribution truncated to [-2, 2] (\cite{feinberg2016conducting, mislevy1989consumer}) and then transformed to the intercept parameters ($c_{k} = -a_{k}b_{k}$). The true item parameter values can be found in the Appendix. The item parameters are treated as fixed across all conditions. Within each replication, the cluster-level LRV, $\theta_{j}$ was generated from $\mathcal{N}(0, 0.5^2)$ (i.e., $\psi=0.5$) and the individual-level deviations, $\delta_{ij}$ was generated from $\mathcal{N}(0, 1)$.


\noindent \textbf{Step 2. Treatment assignment}
Participants are classified based on the ORV. The ORV for the HRD model is the average sum score of item responses within each cluster and that for the MRD model is the individual sum score, as described in Equations \ref{eq:treatment.hrd} and \ref{eq:treatment.mrd}. In practice, the cutoff for the treatment assignment either naturally exists or is chosen based on the research question. In the current study, the cutoff was arbitrarily chosen so that approximately $40\%$ of the data falls below the cutoff ($c=4$). The ORV scores were compared to the cutoff value. Participants were assigned to treatment if their scores were at or below the cutoff and the control group if their scores were above the cutoff. 

\noindent \textbf{Step 3. Generating outcome} 
The outcome was generated following Equations \ref{eq:HRD.comb} and \ref{eq:MRD.comb} for HRD and MRD model, respectively. Three different levels of effect sizes were considered for the ATE given LRV (see Equations \ref{eq:omega.hrd} and \ref{eq:omega.mrd}). We considered regression coefficients for the treatment effect that yield ATE at $40$th percentile LRV approximately equals to $0$, $0.3$, and $0.5$. That gives $\gamma_{01}=0.012, 0.312$, and $0.512$ for the HRD model and $\gamma_{20} = 0.025, 0,325$, and $0.525$ for the MRD model.
\\
The true values for the rest of the structural parameters are as follows: For the HRD model, $\gamma_{00}=0$, $\gamma_{02}=0.30$, $\gamma_{10} = 0.30$, $\gamma_{03}=0.10$, $\tau_{0}=0.5$, and $\sigma=1$. For the MRD model, $\gamma_{00}=0$, $\gamma_{01}=0.10$, $\gamma_{10}=0.30$, $\gamma_{20}=0.10$, $\gamma_{30}=0.10$, $\tau_{0}=0.5$, $\tau_{2}=0.5, \tau_{02}=0$, and $\sigma=1$.

\subsection{Procedure and analysis details} We first fitted the model by using the MH-RM algorithm described in the previous section. The algorithm was implemented in R (\cite{rcore}). The following are the tuning constants involved in the MH-RM algorithm. The gain constants $g_t$ were fixed at $1$ during the burn-in stage and the second stage. For the third stage, $g_t = \frac{1}{t^{\epsilon}}$, where $t$ is the iteration number and $\epsilon=0.51$. $100$ and $500$ iterations were conducted for the burn-in and the second stage, respectively. The maximum number of iterations for the third stage was $1000$, while the convergence was examined by the consecutive change in parameter estimates and maximum gradient. The third stage was deemed converged when the two criteria were below $0.001$. To estimate the standard errors, $2000$ imputations of random effects were conducted after the convergence.
\\
After obtaining the parameter estimates and corresponding standard errors, the ATE conditional on every possible ORV score were computed. The ATE conditional on ORV can be computed as Equations \ref{eq:tau.hrd} and \ref{eq:tau.mrd}. The computation requires estimating the posterior distribution of LVs (i.e., $\theta_j | S(\boldsymbol{X}_j)=s$). Since this computation involves dealing with numerous response patterns, the Lord-Wingersky algorithm (\cite{LordWingersky1984}) was adopted. Finally, the standard error of the ATE was computed using the delta method, as the ATE can be expressed as a differentiable function of the estimated parameters. The entire procedure was replicated $500$ times.

\subsection{Evaluation criteria}
The recovery of the model parameters and ATE given ORV were evaluated in terms of their bias and root mean squared error (RMSE). The true value of the ATE was computed by plugging-in the true measurement and structural parameters. We also check the coverage of the $95\%$ confidence interval of the ATE. The ATE was evaluated at multiple evaluation points within a range of ORV values around the cutoff to confirm that the inference away from the cutoff is valid. 

\subsection{Result}
\begin{figure}[H]
    \begin{center}
	\includegraphics[scale=1]{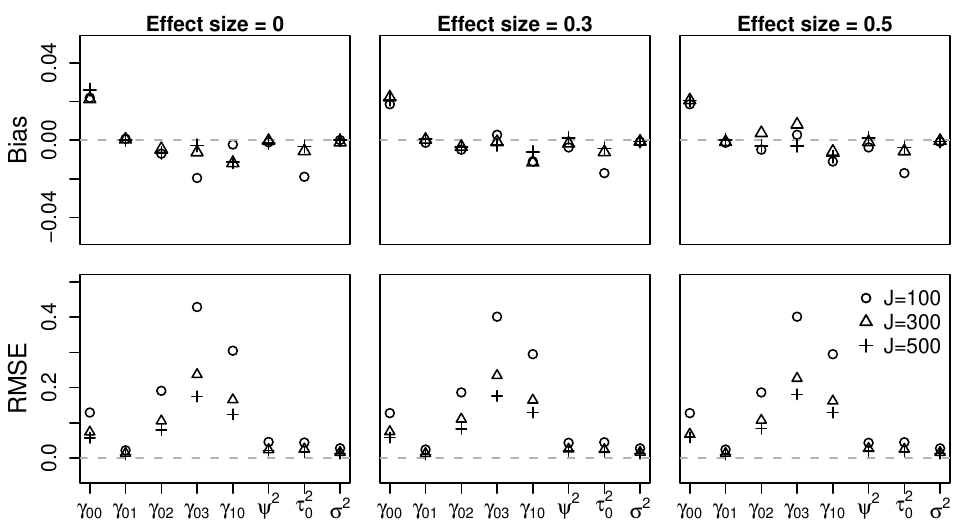}
    \end{center}
\caption{Bias and RMSE of structural parameters for the HRD model. $J:$ the number of clusters.}	
\label{fig:hrd.bias.rmse}
\end{figure}

\begin{figure}[H]
    \begin{center}
	\includegraphics[scale=1]{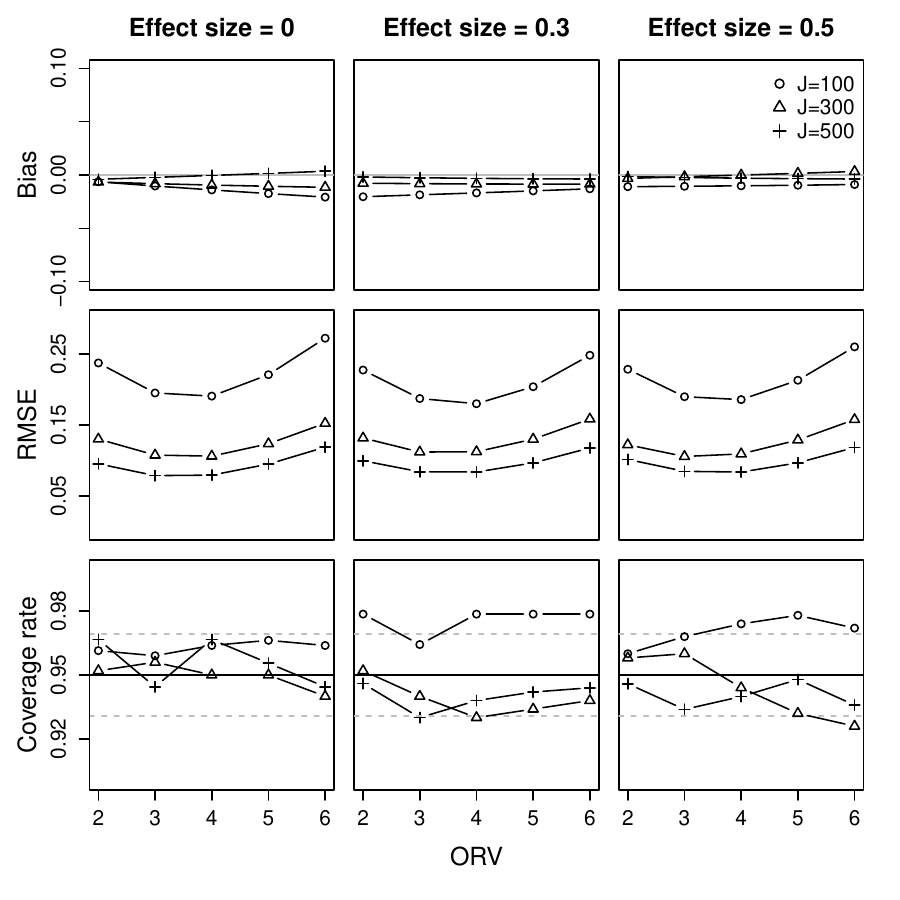}
    \end{center}
\caption{Bias, RMSE and coverage of ATE given ORV values under HRD model. The ORV is the cluster-average summed score. $J:$ the number of clusters. The dotted horizontal lines for the coverage rate indicate normal-approximated $95\%$ Monte-Carlo confidence band.}	
\label{fig:hrd.late}
\end{figure}

\textbf{Hierarchical RD model}.
Figure \ref{fig:hrd.bias.rmse} illustrates the bias and RMSE of the HRD model structural parameters across manipulated conditions. Overall, parameter recovery improved with a larger cluster-level sample size. Variations in effect size had minimal impact on the accuracy of the parameter estimation. Figures \ref{fig:hrd.late} shows the bias, RMSE and coverage rate of the ATE given ORV scores. For the HRD model, ORV values $-1, -0.5, +0.5,$ and $ +1$ away from the cutoff were investigated. While a wider range of ORV values could be explored, the $c\pm1$ is already considered a relatively broad range for the extrapolation in the HRD model, as the ORV represents cluster-average summed score. Notably, the standard deviation of cluster-average summed scores is around $1$ and more than $60\%$ of the total clusters fall within the $c\pm 1$ range.
\\
The first and second rows of Figure \ref{fig:hrd.late} shows bias and RMSE of the ATE, respectively. The bias remained minimal across manipulated conditions and across ORV values. On the other hand, the RMSE increased as ORV deviated further from the cutoff, suggesting that variability increases when the ATEs are estimated away from the cutoff. Also, RMSE was larger when smaller cluster-level sample size was adopted.
\\
The third row of Figure \ref{fig:hrd.late} illustrates the coverage of the 95\% confidence interval for the ATE. The dotted horizontal lines represent the normal-approximated 95\% Monte Carlo confidence interval based on 500 replications. The coverage is considered accurate if it falls within this band, accounting for the Monte Carlo error. With 100 clusters, slight over-coverage appears for some ORV values, especially for the effect size of $0.3$ and $0.5$. However, as the number of clusters increases, coverage improves, ultimately falling within the confidence band across all ORV values with $500$ clusters. The result implies that the estimation of the ATE is accurate across ORV levels when sufficiently large number of clusters are available, although the efficiency of the estimation may decrease when ORV moves away from the cutoff.

\begin{figure}[H]
    \begin{center}
	\includegraphics[scale=1]{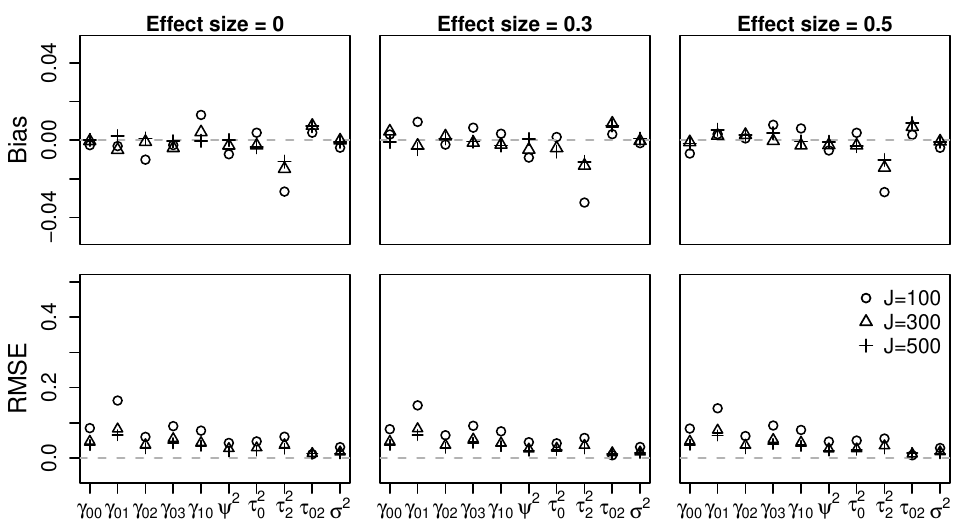}
    \end{center}
\caption{Bias and RMSE of structural parameters for the MRD model. $J:$ the number of clusters.}	
\label{fig:mrd.bias.rmse}
\end{figure}

\begin{figure}[H]
    \begin{center}
	\includegraphics[scale=1]{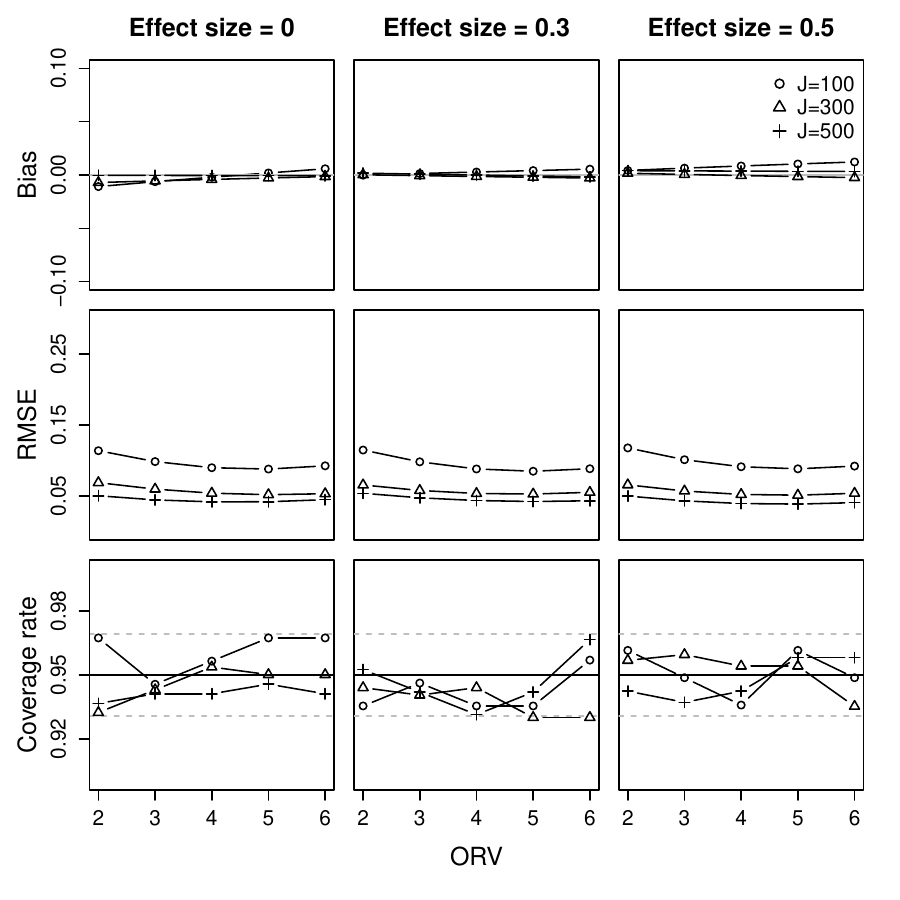}
    \end{center}
\caption{Bias, RMSE and coverage of ATE given ORV values under MRD model. The ORV is the cluster-average summed score. $J:$ the number of clusters. The dotted horizontal lines for the coverage rate indicate normal-approximated $95\%$ Monte-Carlo confidence band.}
\label{fig:mrd.late}
\end{figure}

\noindent \textbf{Multisite RD model}.
Figure \ref{fig:mrd.bias.rmse} illustrates the bias and RMSE of the MRD model structural parameters. Recovery of the parameters was better for the MRD model compared to the HRD model. The RMSEs of the parameter estimates were much smaller than those from the HRD model. This could be because the number of units that contribute the estimation is larger for some parameters as the treatment assignment is determined at the individual level.
\\
Figure \ref{fig:mrd.late} presents the bias, RMSE and coverage rate of the ATE given ORV values. For the MRD model, ORV range $c \pm 2$ was investigated, which includes approximately $70\%$ of the total individuals. We observe a similar pattern to that observed in the HRD model, indicating that the ATEs are accurately estimated across ORV values. Again, the estimation becomes less efficient as the evaluation point moves further from the original cutoff. The bias and RMSE were in general smaller compared to those from the HRD model. In addition, the coverage of the ATE fell within the confidence band across all ORV values even with the smallest sample size. No obvious discrepancies were observed across different effect sizes.

\section{Discussion}
This work extends the latent RD framework to multilevel RD designs, enabling useful augmentations such as extrapolating ATEs and quantifying heterogeneous treatment effects for multilevel RD analysis. We accommodate two common multilevel RD designs—Hierarchical and Multisite RD—each suited to different scenarios depending on whether the assignment occurs at the cluster level or the individual level. Our simulation study demonstrates that the proposed models can be accurately fitted given a sufficiently large cluster-level sample size. ATEs are also well-estimated at ORV values beyond the original cutoff score. The framework allows for estimating the cluster-level ATE using the HRD model or the individual-level ATE using the MRD model when a cutoff different from the original is applied. Additionally, it enables the quantification of ATE variability among clusters (in the HRD model) or individuals (in the MRD model) with the same ORV score.
\\
The estimation efficiency may decrease when ATEs are extrapolated too far away from the original cutoff. We clarify that the range of extrapolation we examined is relatively broad. In practice, extrapolating ORVs too far from the original cutoff requires caution due to both theoretical and practical reasons. The rationale behind the extrapolation is that for the entire range of LRV values, there exists a positive probability for an individual to be assigned to either treatment or control group, due to the measurement error. Still, this probability will decrease as the individual's latent ability is located far away from the cutoff score. In addition, extrapolating too far from the cutoff may not be practically meaningful. For example, if an intervention is designed specifically for gifted students and the original cutoff is set accordingly, estimating the ATE at significantly lower ORV levels would contradict the intent of the intervention and yield results that lack practical relevance.
\\
We acknowledge that the current work has limitations and can be extended in the following directions.
\\
First, this study limited the structural models to be essentially the same as the original suggestion by \textcite{RhoadsDye2016} with observed RV. However, researchers could incorporate additional random effects based on their research interest. Additionally, the models examined in this study did not incorporate covariates. However, they can be extended to include observed and/or latent covariates. Although the RD design does not require covariates to identify causal treatment effects, researchers often include them to enhance the robustness of the analysis. When a covariate is latent, an additional measurement model can be jointly estimated.
\\
Second, we assumed a simple unidimensional latent structure underlying the LRV. However, more complex latent structures, such as multidimensional measurement models, can also be specified for the LRV. Moreover, in practice, the eligibility criteria can be multiple rather than single. The identification of treatment effects in such cases has been discussed by \textcite{wong2013analyzing}. This multivariate assignment could even be modeled using multidimensional measurement models under the latent RD framework. The corresponding identification of treatment effects requires further investigation.
\\
Third, extensive simulation work could be conducted to further investigate the performance of the proposed model under varying conditions. The current study assumed that cluster sizes were balanced, the ICC was fixed at $0.25$ and test length was fixed at $10$. However, in real-world applications, cluster sizes often vary, and ICC values may differ depending on the context. Also, the test length impacts the amount of measurement error that is directly associated with the variability of the LV posterior, which is a key quantity in this framework. Future research could explore the impact of variations in these factors on model estimation and inference.

\clearpage
\section{Appendix}

\subsection{A1. Recovery of measurement parameters}
\begin{figure}[H]
    \begin{center}
	\includegraphics[scale=1]{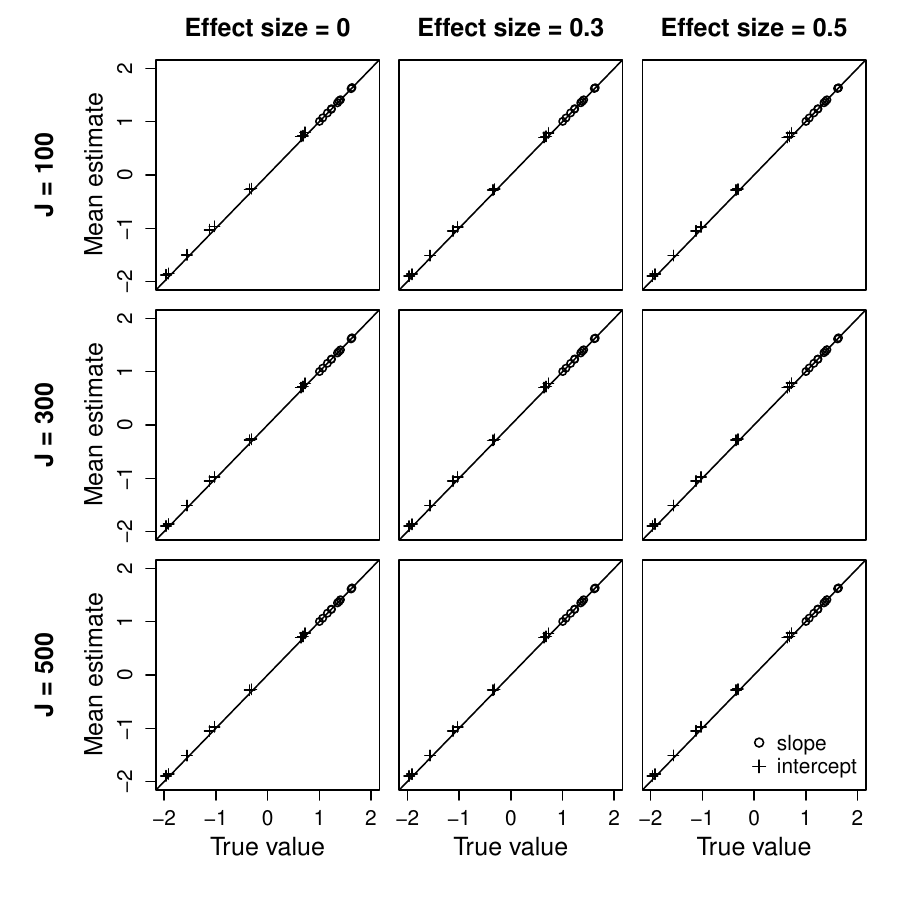}
    \end{center}
\caption{Mean estimate of measurement parameters plotted against true values under HRD model.}	
\label{fig:hrd.measurement.recovery}
\end{figure}

\begin{figure}[H]
    \begin{center}
	\includegraphics[scale=1]{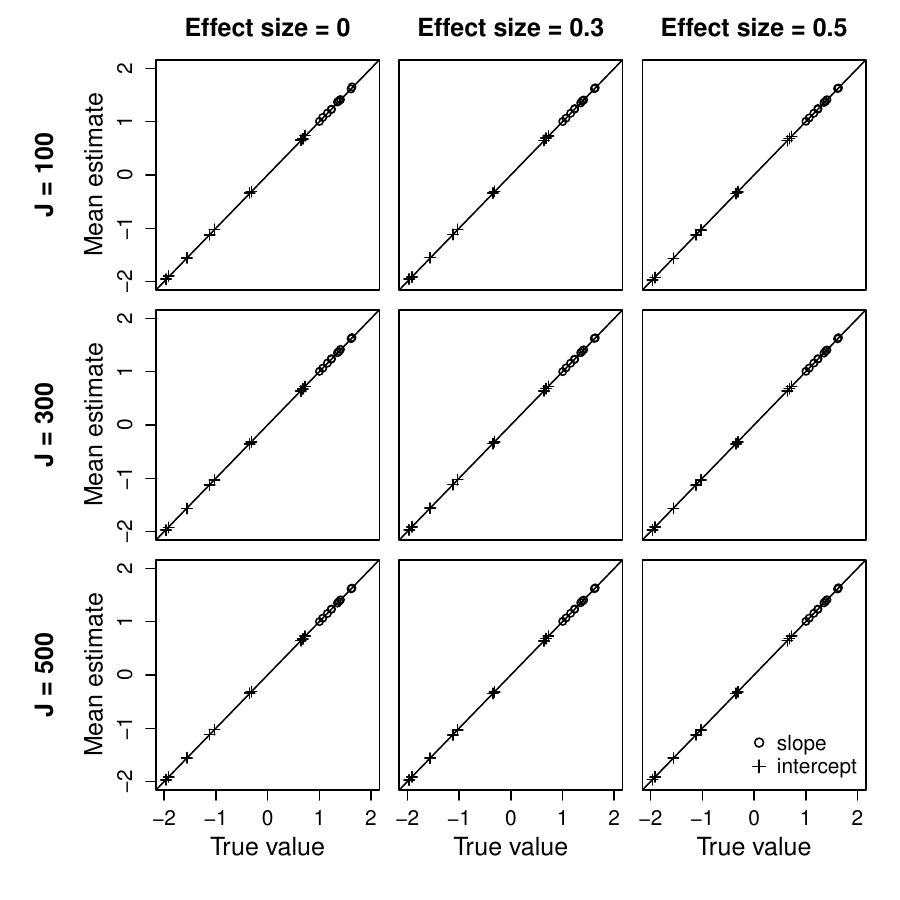}
    \end{center}
\caption{Mean estimate of measurement parameters plotted against true values under MRD model.}	
\label{fig:hrd.measurement.recovery}
\end{figure}

\subsection{A2. Derivatives of the complete data log-likelihood}
\renewcommand{\theequation}{A\arabic{equation}}
\setcounter{equation}{0}  

The complete-data likelihood function of the HRD model can be written as
\begin{equation}
  \prod_{j = 1}^J\left\{ 
    \prod_{i = 1}^{I_j}\left[ 
      \prod_{k = 1}^K
      f_{x|\delta, \theta}(x_{ijk}|\delta_{ij}, \theta_j)
    \right]
    f_{y|\delta,\theta,u}(y_{ij}|\delta_{ij}, \theta_j, u_{0j})
    f_\delta(\delta_{ij})
  \right\}
  f_\theta(\theta_j)f_u(u_{0j}).
  \label{eq:hrd.comp.lik}
\end{equation}

\noindent The derivatives with respect to the item parameters are only associated to $\prod_{k = 1}^K f_{x|\delta, \theta}(x_{ijk}|\delta_{ij}, \theta_j)$. Recall the multilevel 2-PL model (Equation \ref{eq:mlikx}). Suppressing subscripts for brevity, define $P$ as
\begin{align}
    P = \frac{\exp[a(\theta+\delta)+c]}{1+\exp[a(\theta+\delta)+c]}
    \label{eq:P}
\end{align}
The log-likelihood of interest can be written as
\begin{align}
    l = \sum_{k=1}^{K} \left[ x \log P + (1-x) \log (1-P) \right]
    \label{eq:meas.loglik}
\end{align}
The first derivatives are as follows:
\begin{align*}
    \frac{\partial l}{\partial a} =&\ \frac{\partial l}{\partial P} \frac{\partial P}{\partial a}, \cr
    \frac{\partial l}{\partial c} =&\ \frac{\partial l}{\partial P} \frac{\partial P}{\partial c},
    \label{eq:meas.first.deriv}
\end{align*}
where
\begin{align}
    \frac{\partial l}{\partial P} =&\ \frac{x-P}{P(1-P)}, \cr
    \frac{\partial P}{\partial a} =&\ P(1-P)(\theta + \delta), \cr
    \frac{\partial P}{\partial c} =&\ P(1-P).
\end{align}
The second derivatives are given by following:
\begin{align*}
    \frac{\partial^2 l}{\partial a \partial a} =&\ \frac{\partial^2 l}{\partial a \partial P} \frac{\partial P}{\partial a} + \frac{\partial l}{\partial P} \frac{\partial^2 P}{\partial a \partial a}, \cr
    \frac{\partial^2 l}{\partial c \partial c} =&\ \frac{\partial^2 l}{\partial c \partial P} \frac{\partial P}{\partial c} + \frac{\partial l}{\partial P} \frac{\partial^2 P}{\partial c \partial c}, \cr
    \frac{\partial^2 l}{\partial a \partial c} =&\ \frac{\partial^2 l}{\partial a \partial P} \frac{\partial P}{\partial c} + \frac{\partial l}{\partial P} \frac{\partial^2 P}{\partial a \partial c},    
\end{align*}
where
\begin{align}
    \frac{\partial^2 l}{\partial a \partial P} =&\ \left( -\frac{x}{P^2} - \frac{1-x}{(1-P)^2}  \right) \frac{\partial P}{\partial a}, \cr
    \frac{\partial^2 l}{\partial c \partial P} =&\ \left( -\frac{x}{P^2} - \frac{1-x}{(1-P)^2}  \right) \frac{\partial P}{\partial c}, \cr
    \frac{\partial^2 P}{\partial a \partial a} =&\ P(1-P)(1-2P)(\theta + \delta)^2, \cr
    \frac{\partial^2 P}{\partial c \partial c} =&\ P(1-P)(1-2P), \cr
    \frac{\partial^2 P}{\partial a \partial c} =&\ P(1-P)(1-2P)(\theta + \delta).
\label{eq:meas.second.deriv}
\end{align}

The derivatives with respect to regression coefficients and residual variance $\sigma^2$ are only associated to $f_{y|\delta,\theta,u}(y_{ij}|\delta_{ij}, \theta_j, u_{0j})$. The log-likelihood can be written as
\begin{align}
    l= -\frac{N}{2} \log{2\pi\sigma^2} - \sum_{i=1}^{I_{j}}\sum_{j=1}^{J} \frac{1}{2\sigma^2}(y - \mu)^2,
    \label{eq:str.loglik}
\end{align}
where $N = J \cdot I_{j}$ and we define $\mu$ as $\gamma_{00} - \gamma_{01}\theta_{j} - \gamma_{02}g - \gamma_{03} \theta_{j} g  - \gamma_{10}\delta_{ij} - u_{0j}$. The first derivative with respect to a specific regression coefficient $\gamma_{f}$ is 
\begin{align}
    \frac{\partial l}{\partial \gamma_{f}} = \frac{1}{\sigma^2} \frac{\partial \mu}{\partial \gamma_{f}} (y - \mu)
    \label{eq:str.first.deriv}
\end{align}
The second derivative with respect to regression coefficients $\gamma_{f}$ and $\gamma_{f'}$ is
\begin{align}
    \frac{\partial^2 l}{\partial \gamma_{f} \gamma_{f'}} = \frac{1}{\sigma^2} \frac{\partial \mu}{\partial \gamma_{f}} \frac{\partial \mu}{\partial \gamma_{f'}}
\label{str.second.deriv}
\end{align}
The first derivative with respect to variance parameter $\sigma$ is
\begin{align}
    \frac{\partial l}{\partial \sigma} = -\frac{N}{2}\sigma^{-2} + \sum_{i=1}^{I_{j}}\sum_{j=1}^{J}(y-\mu)^2 \sigma^{-3}
\label{eq:sigma.first.deriv}
\end{align}
The second derivative with respect to variance parameter $\sigma$ is
\begin{align}
    \frac{\partial^2 l}{\partial \sigma \partial \sigma} = N \sigma^{-3} - 3\sigma^{-4}(y-\mu)^2
\label{eq:sigma.second.deriv}
\end{align}
\\
The complete-data likelihood function of the MRD model can be written as
\begin{equation}
  \prod_{j = 1}^J\left\{ 
    \prod_{i = 1}^{I_j}\left[ 
      \prod_{k = 1}^K
      f_{x|\delta, \theta}(x_{ijk}|\delta_{ij}, \theta_j)
    \right]
    f_{y|\delta,\theta,\boldsymbol{u}}(y_{ij}|\delta_{ij}, \theta_j, u_{0j}, u_{2j})
    f_\delta(\delta_{ij})
  \right\}
  f_\theta(\theta_j)f_{\boldsymbol{u}}(u_{0j}, u_{2j}).
  \label{eq:mrd.comp.lik}
\end{equation}
The HRD and MRD model shares the same measurement model and therefore, the derivatives with respect to the item parameters are identical. Also, the derivatives with respect to the regression coefficients and residual variance $\sigma^2$ are essentially the same as Equations \ref{eq:str.first.deriv} - \ref{eq:sigma.second.deriv}, where $\mu$ is now defined as $\gamma_{00} - \gamma_{01}\theta_{j} - \gamma_{10}\delta_{ij} - (\gamma_{20} + u_{2j})g - $ $\gamma_{30} \delta_{ij} g$ $- u_{0j}$.
\\
The derivatives with respect to the variances of random effects are associated with $f_{\boldsymbol{u}}(u_{0j}, u_{2j})$, which is a bivariate normal distribution. The log-likelihood is
\begin{align}
    l &= -\frac{N}{2}\log 2\pi - \frac{1}{2}\log |\boldsymbol{\Sigma}| - \boldsymbol{u}\t\boldsymbol{\Sigma}^{-1}\boldsymbol{u}, \cr
    &\text{where} \,\, \boldsymbol{u} = (u_{0j}, u_{2j})\t \text{and}\; \SSigma= \begin{pmatrix}
	\tau_0^2 & \tau_{02}\\
	\tau_{02} & \tau_2^2\\
	\end{pmatrix}
\end{align}
The first derivatives with respect to the variance $\SSigma$ is
\begin{align}
    \frac{\partial l}{\partial \SSigma} &= \frac{1}{2}vecs(\boldsymbol{u}\boldsymbol{u}\t - \SSigma)\t\boldsymbol{D}\t(\SSigma^{-1} \otimes \SSigma^{-1}) \boldsymbol{D} \cr
    &\text{where $\boldsymbol{D}$ is the duplication matrix.} 
\end{align}
The second derivatives with respect to the variance $\SSigma$ is
\begin{align}
    \frac{\partial^2 l}{\partial \SSigma \partial \SSigma} &= \frac{1}{2}\boldsymbol{D}\t[\SSigma^{-1}\otimes\{ \SSigma^{-1} (2\boldsymbol{W}-\SSigma)\SSigma^{-1} \}]\boldsymbol{D}, \cr
    &\text{where $\boldsymbol{W} = \boldsymbol{S} + \bar{\boldsymbol{u}} \bar{\boldsymbol{u}}\t$ }.
\end{align}
$\boldsymbol{S}$ denotes the sample covariance matrix and $\bar{\boldsymbol{u}}$ denotes the vector of sample means.

\printbibliography

\end{document}